\documentclass[nonatbib,preprint,proceedings]{rmaa}

\usepackage{natbib}

\SetYear{2006}
\SetConfTitle{Circumstellar Media and Late Stages of Massive Stellar Evolution}

\title{Evidence for Pre-SN Mass Loss in the Galactic SNR 3C~58}

\author{
  Gwen C. Rudie\altaffilmark{1} 
  and Robert A. Fesen\altaffilmark{1}}

\altaffiltext{1}{Department of Physics \& Astronomy, Dartmouth College, Hanover, NH, USA.}

\shortauthor{Rudie \& Fesen}
\shorttitle{Progenitor Mass Loss in 3C~58}

\fulladdresses{
\item Gwen C. Rudie and Robert A. Fesen: 6127 Wilder Laboratory, Physics and Astronomy Department,
   Dartmouth College, Hanover, NH 03755, USA
  (\email{rudie@dartmouth.edu, fesen@snr.dartmouth.edu}).
}

\listofauthors{Gwen C. Rudie \& Robert A. Fesen}
\indexauthor{Rudie, G. C.}
\indexauthor{Fesen, R. A.}

\abstract{We discuss the findings of a comprehensive imaging and spectroscopic
survey of the optical emission associated with the supernova remnant 
3C~58 \citep{fes06} as they relate to the topic of pre-SN mass loss.
Spectroscopically measured radial velocities of $\sim$450 emission knots within
the remnant show two distinct kinematic populations of optical knots: a
high-velocity group with radial velocities in the range of 700 -- 1100 km~s$^{-1}$ and
a lower velocity group exhibiting radial expansion velocities below $\sim$250
km~s$^{-1}$. We interpret the high-velocity knots as ejecta from the SN explosion and
the low-velocity knots as shocked circumstellar material likely resulting from
pre-SN mass loss. The chemical signatures of the two populations also show
marked differences. The high velocity group includes a substantial number of
knots with notably higher [\ion{N}{2}]/H$\alpha$ ratios not seen in the lower
velocity population, suggesting greater nitrogen enrichment in the SN ejecta
than in the CSM.  These results are compared with evidence for pre-SN mass loss in
the Crab Nebula, perhaps the SNR most similar to 3C~58. These SNRs
may comprise two case studies of pre-SN mass loss in relatively low mass ($\sim$8
-- 10 $M_{\odot}$) core-collapse SN progenitors.}

\addkeyword{ISM: circumstellar material}
\addkeyword{ISM: individual (3C~58)}
\addkeyword{ISM: individual (Crab Nebula)}
\addkeyword{ISM: supernova remnants}

\begin{document}
\maketitle

\section{Introduction}
\label{sec:intro}
\begin{figure*}[!t]\centering
  %
  \includegraphics[width=\textwidth]{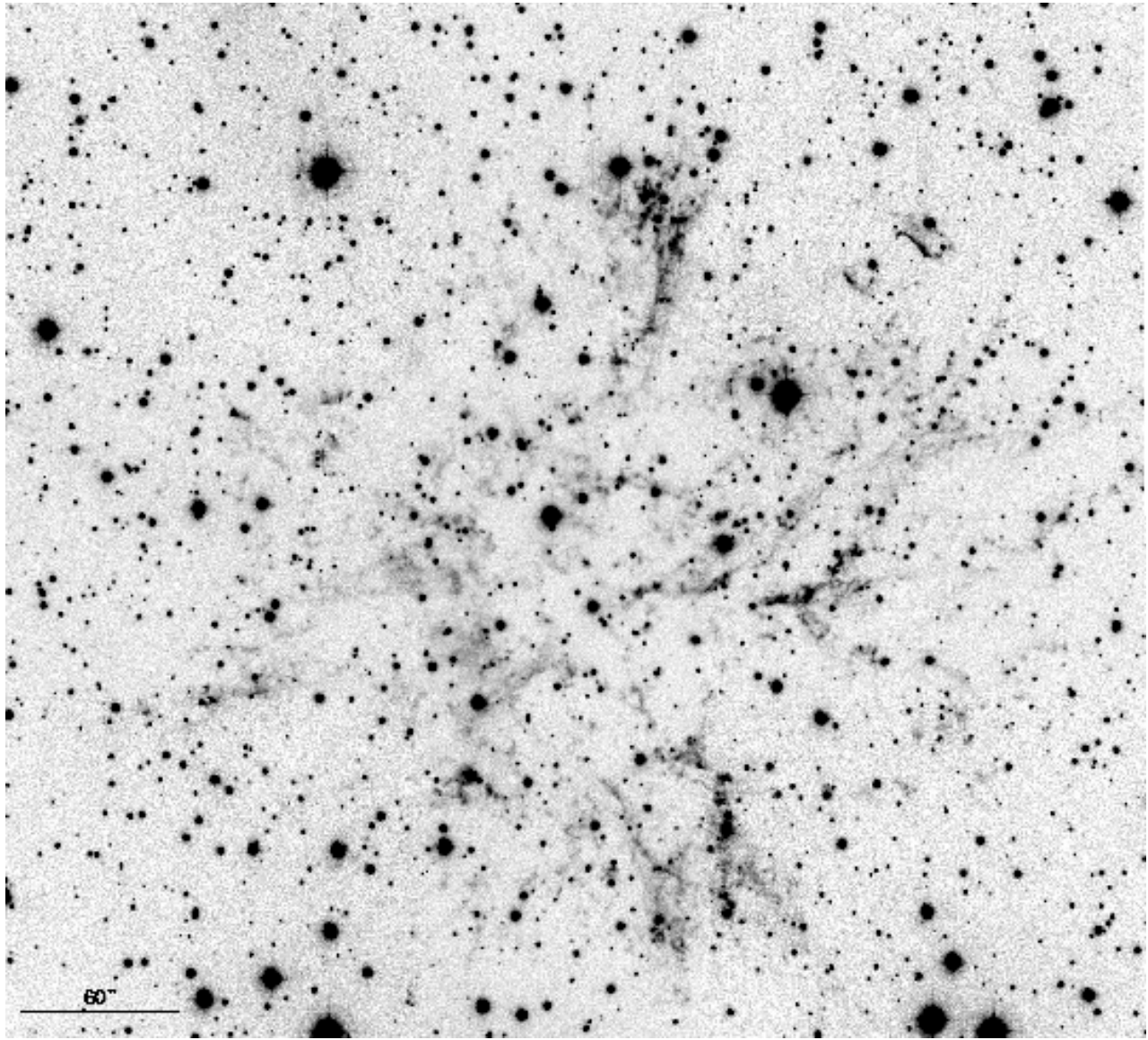}
\caption{A deep H$\alpha$ image of the central regions of the 3C~58 supernova remnant \citep{fes06}.}
\label{fig:Hadeep}
\end{figure*}

3C~58 is a Galactic supernova remnant (SNR) with many similar properties to the
Crab Nebula. Both remnants have rapidly spinning neutron stars (pulsars) which strongly
influence their radio and X-ray morphologies. 'Crab-like' or `plerionic'
(filled-center) remnants, like 3C~58, are brightest toward their centers in
both X-rays and in the radio \citep{WS71,WW76}. The bright central emission in
plerions is a result of the interaction between strong magnetic fields
and the outflow of relativistic particles from the neutron star. This interaction
produces the synchrotron emission seen in the resulting pulsar wind nebulae (PWN).

Aside from their classification as plerionic remnants, 3C~58 and the Crab share
a number of other properties. Both are thought to have progenitors in the
initial main sequence mass range of 8 -- 10 $M_{\odot}$
\citep{nom82,nom85,nom87,Fesen83}.  Both remnants are also believed to be
relatively young, with the Crab being the well established remnant of the
historic guest star of 1054, while 3C~58 is proposed to be the SNR of a 
guest star seen in 1181 \citep{S71,CS77,SG02}.  The two remnants also exhibit similar
expansion velocities: 3C~58 and the Crab have maximum $V_R$ of 1100 km~s$^{-1}$ and
2200 km~s$^{-1}$ \citep{Clark83,FK85,Law95} respectively. Such expansion velocities
are comparatively low for young SNRs implying a much lower kinetic energy of
approximately $10^{49.5}$ erg compared with the canonical $10^{51}$ erg.

However, despite their similar expansion velocities and possibly similar ages,
the remnants are quite different in physical size.  The Crab nebula is located
at a distance of $\sim$2 kpc \citep{DF85} and has an angular size of roughly 5'
x 7' \citep{Wilson72} making it  3 x 4 pc in size. 3C~58 is farther away
at $\sim$3 kpc \citep{GG82,Roberts93} and roughly 6' x 10' (see Fig.\ 1)
\citep{WW76,RA88}, making it 5 x 9 pc or around twice the size of the Crab.

\section{Evidence for Progenitor Mass Loss in 3C~58}
\label{sec:csm }

In a recent comprehensive spectral survey of 3C~58 completed by \citet{fes06},
radial velocities and compositions of 463 emission knots within the remnant
were measured. The observed radial velocities in 3C~58 range from --1070 km~s$^{-1}$
to +1100 km~s$^{-1}$ and form two kinematically distinct velocity groups as shown in
Fig.~2. The high velocity group centers around 770 km~s$^{-1}$ and forms a thick shell
with a velocity dispersion of $\pm155$ km~s$^{-1}$. The second group consists of lower
velocity emission knots. Measured radial velocities for this group range
between --250 km~s$^{-1}$ and +200 km~s$^{-1}$. There is a clear separation between these
populations which only merge at radial distances from the pulsar of
$\sim$100$''$. Both groups of emission knots show a fairly uniform spatial
distribution across the remnant. 

\begin{figure}
\includegraphics[angle=270,width=0.45\textwidth]{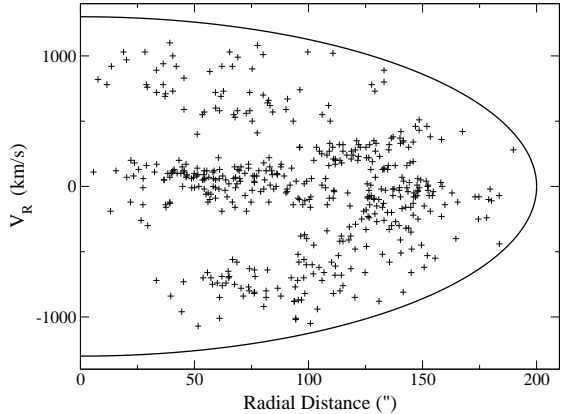}
\caption{Plot of observed knot radial velocity with projected knot radial distance from 
the remnant's central X-ray point source
(PSR J0205+6449). The curve in the figure represents a spherical expansion of 1300 km~s$^{-1}$.}
\label{fig:vel_dist}
\end{figure}

We interpret the presence of these two distinct populations as evidence for
both SN ejecta and circumstellar material (CSM) within the 3C~58 remnant. The
higher radial velocity knots are likely the ejecta from the SN explosion
itself, while the slower ones may be shocked CSM. We believe these CSM knots to
be the result of a pre-SN mass loss phase during which material from the
progenitor star was ejected with velocities of $0 - 200$ km~s$^{-1}$. The low-velocity
population of suspected CSM emission knots is extensive, making up
approximately half of the optically emitting material near the remnant's
center. The lack of intermediate velocity knots within the remnant (see Fig.\
2) suggest the high-velocity ejecta, like that seen in the Crab Nebula, is
largely confined to an outer shell. 

The distribution of knot density in 3C~58 is compared to radial velocity in Figure 3.
Because the [\ion{S}{2}] emission measurements from the fainter knots have
significant errors, the plot only reflects the ratios of the brightest of
3C~58's knots. In this figure, the circle size corresponds to the [\ion{S}{2}]
6716/6731 line ratio which is a density sensitive ratio. Higher [\ion{S}{2}]
6716/6731 line ratios (larger circles) are representative of lower densities.
The two velocity populations show no correlation to any sort of density pattern
within the remnant.

\begin{figure}
\includegraphics[angle=270,width=0.45\textwidth]{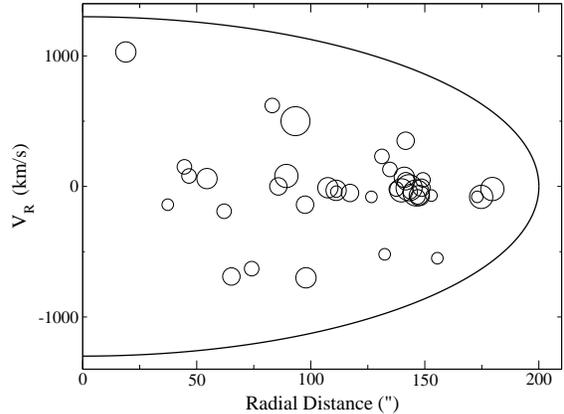}
\caption{Radial velocity of knots in 3C~58 with H$\alpha$ flux 
$\geq$ $1 \times 10^{-15}$ erg cm$^{-2}$ s$^{-1}$.
Symbol size is proportional to the measured [\ion{S}{2}] 6716/6731 line ratio; 
hence, the smaller symbol size indicates
higher knot electron density.}   
\label{fig:density_plot}
\end{figure}

In terms of composition, the two kinematic populations of emission knots differ
noticeably.  Since the [\ion{N}{2}] 6548,6583 lines are not major nebula
coolants, the observed [\ion{N}{2}]/H$\alpha$ ratio can be used as a
first order indicator of relative N/H abundances.  Within the high-velocity
group, there is a substantial number of knots which show relatively high
[\ion{N}{2}]/H$\alpha$ ratios. This trend is shown in Figure~4 which plots the
correlation between the velocity pattern seen in the remnant and the
composition of these two velocity groups.  The circle size in this figure
represents the observed [\ion{N}{2}]/H$\alpha$ line ratio. Knots with
intermediate and low [\ion{N}{2}]/H$\alpha$ ratios are found in both the high
and low velocity populations.  However, knots showing relatively high
[\ion{N}{2}]/H$\alpha$ ratios (i.e., the larger circles) are confined to the
high-velocity knot population.

Interestingly, the highest [\ion{N}{2}]/H$\alpha$ knots lie toward the middle
or inner edge of the maximum velocity range while the highest velocity knots
show values closer to the average  [\ion{N}{2}]/H$\alpha$ ratio. We would expect the
highest velocity ejecta to originate from the surface of the star (thus having
a composition similar to the CSM) while somewhat slower SN ejecta would have
come from deeper within the star and therefore might exhibit a greater degree
of nitrogen enrichment. This is the basic structure of the composition pattern observed
within 3C~58. Thus, the composition measurements from the remnant support the
scenario in which the low-velocity material is circumstellar mass loss ejected
by the progenitor prior to the SN event, while the high-velocity material
represents SN ejecta.

In general, the measured chemical abundances of the emission knots within the 
remnant support the notion of pre-SN mass loss. 3C~58 commonly exhibits
strong [\ion{N}{2}] emission ([\ion{N}{2}]/H$\alpha$ $\simeq 3 - 10$) which is
considerably higher than ratios seen in shocked gas with solar composition
\citep{CoxRaymond85,Hartigan94}. These ratios indicated nitrogen enrichment at
least several times solar \citep{MacAlpine96,ZK00}, likely caused by CNO
processing within the progenitor star. Similar [\ion{N}{2}] line emission
ratios have been observed in mass loss material found in Wolf-Rayet nebulae
such as NGC 6888 \citep{Kwitter81} and in young SNRs such as the QSFs in Cas A
\citep{vdb71,Pvdb71,Fesen01}. This supports the conclusion that the
low-velocity population is nitrogen-enriched stellar material shed by the progenitor
before the SN event which has since been shocked by the expanding SNR.

\section{Similarities between 3C~58 and the Crab Nebula}

In terms of its radio and X-ray emission, 3C~58 has long been viewed as a
remnant with properties most similar to those of the Crab Nebula. However, it now
appears that the 3C~58 and Crab Nebula remnants may have an added similarity as
both remnants show evidence for the presence of pre-SN mass loss.

The chief arguments in favor of CSM in the Crab Nebula lie in the chemical
signature of an east-west band of He--rich filaments that circle the center of
the remnant \citep{MacAlpine89} and the synchrotron `bays' on the eastern and
western edges of these filaments \citep{Fesen92}.  Additional evidence comes
from several outlying filaments in the Crab which emit strong [\ion{N}{2}]
emissions compared to H$\alpha$ \citep{FK82,MacAlpine96}, not unlike that seen
in the CSM knots in 3C~58.

\begin{figure}
\includegraphics[angle=270,width=0.45\textwidth]{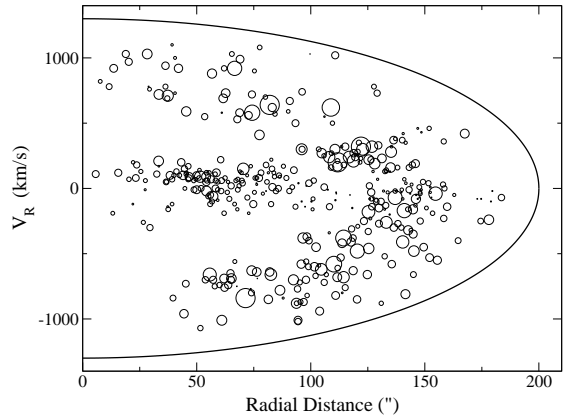}
\caption{Radial velocity of knots in 3C~58. Symbol size is proportional 
to the measured [\ion{N}{2}]/H$\alpha$ line ratio.}   
\label{fig:nitrogen_plot}
\end{figure}

On the other hand, the distribution of pre-SN mass loss material in the two
remnants appears to be dissimilar. In 3C~58, the low-velocity CSM knots are
uniformly distributed across the remnant. In contrast, within the Crab Nebula,  
the circumstellar material was likely confined to a toroidal disk. The remnant of this
disk is observed today as a helium--rich strip running east to west across the
remnant \citep{Uom87} which exhibits a \ion{He}{1} $\lambda5876/H\beta$ line
ratio $>$ 0.9 compared to ratios as low as 0.38 away from this region near the
base of the northern filamentary jet \citep{MacAlpine89}. 

Drift scans of the Crab nebula, in which the slit is oriented N-S, show a 
strong asymmetric N--S velocity pattern.  That is, around the high-He band
location, radial velocities are significantly lower than in the rest of the
remnant. The overall velocity structure of the remnant resembles a figure eight
or an hourglass, with the north and south regions exhibiting bubble-like bipolar
structures. This bipolar expansion pattern might have been caused by SN ejecta confinment
and deceleration by a toroidal disk composed of the high-He
filaments \citep{MacAlpine89,Fesen97}. Fabry-Perot imaging
spectroscopy of the remnant support such a toroidal morphology of the high--He
band \citep{Law95}. 

The coincidence between the pinched velocity structure and high He filaments
suggest the possibility of a pre-SN structure built up by mass loss from the
progenitor star. The observed He enrichment of this structure may have come from CNO cycle processing of
stellar material prior to its ejection \citep{FK82,MacAlpine89,MacAlpine96}.
Similarly, in 3C~58, the high [\ion{N}{2}]/H$\alpha$ ratios suggest that the
slow moving velocity group is pre-SN mass loss material in which strong He
emission lines are too weak to easily detect.

A confined explosion in the Crab Nebula caused by the presence of an east--west
circumstellar disk might also help explain the presence of the remnant's
synchrotron `bays'; two indentations on the eastern and western edges of the
remnant which contain little synchrotron emission  \citep{Fesen92}.  Optical
proper motion measurements of these bays show them to be moving outward at a
somewhat slower expansion rate, consistent with the constrained velocities in
this region reported by \citet{MacAlpine89}. Polarization studies of the region
also show the edges of the bays have extremely coherent polarization vectors
suggesting well organized magnetic field loops parallel to the edges of the
bays. 

In the pre-SN mass loss model, the Crab's synchrotron bays could be the result
of the magnetic field of the pulsar wrapping around a red-giant or asymptotic
giant branch pre-SN mass loss circumstellar disk, thus blocking the progress of
the charged particles emitting the synchrotron radiation \citep{Fesen92}. The
remains of this circumstellar disk are the high-He band of filaments, and in this
scenario, the magnetic torus is anchored to this band \citep{Fesen92,Smith03}.
\citet{Fesen92} hypothesize the progenitor mass loss could have been induced by
a binary companion which would help confine the CSM to a thin disk capable of
surviving the SN explosion.

In the case of 3C~58, the pre-SN mass loss distribution appears to be quite
different. There is a clear separation between the two velocity populations (see
Fig. 2) suggesting that pre-SN circumstellar mass loss material did not confine the SN
expansion. If the SN ejecta in 3C~58 had been decelerated by CSM present
at the time of explosion, we would expect the confinement to be uneven, leaving
a full range of velocities with no real separation between the two populations.
Further, the velocity pattern observed in 3C~58 does not support the presence of
a circumstellar disk. 

\section{The 3C~58 Progenitor}

In view of the similarities between the Crab and 3C~58 as outlined above, it is
likely that the progenitors of 3C58 and the Crab were similar in nature. The Crab Nebula
has been suggested to be the result of an electron capture SN \citep{nom87}. In
this model, the Crab progenitor evolved into a helium star (with an extended
red-giant-like envelope) via mass loss.
During this helium star phase, electron captures by $^{24}$Mg and $^{20}$Ne
effectively reduce the Chandrasekhar mass of the core, causing the progenitor's
O+Ne+Mg degenerate core to collapse. This would occur prior to oxygen burning
within the core as the smaller mass of this star would be unable to reach the
necessary core temperature \citep{Nomoto84}. 

Electron capture SNe, which are thought to be the end stage of intermediate
mass stars (8 -- 12 $M_{\odot}$), are likely to have lower kinetic energies.
Because these SNe collapse with timescales determined by the electron capture
rate rather than dynamical timescales, they release less kinetic energy. Oxygen
deflagration further acts to decelerate the collapse velocities \citep{nom87}.
Thus, the shock wave in these events is considerably weaker than in canonical
iron core collapse events. If the 3C~58 SN was also an electron capture event
this might help explain its relatively low expansion velocity.

As discussed during this conference, the discovery of the red supergiant (RSG)
progenitor of SN 2003gd \citep{vandyk03,Smartt04,hend05} implies a firm empirical
RSG -- Type IIP SN connection.  This also indicates that low mass ($\sim$8 -- 10
$M_{\odot}$) RSG progenitors may lead to SNe IIP.  While the progenitors of the
Crab and 3C~58 are believed to have been in this mass range, it is unclear if
these two relatively unusual remnants resulted from typical SN~IIP explosions.

\section{Conclusions}

The kinematic and chemical properties of 3C~58's optical emission knots
strongly suggest the existence of appreciable pre-SN circumstellar material
within the remnant. 

3C~58 and the Crab Nebula share many properties which may include the
presence of pre-SN mass loss. Because of these similarities, and the
intermediate mass of these remnant's progenitors, they likely represent the result
of relatively low energy core-collapse SNe brought on by electron capture.  

Finally, if 3C~58 is really associated with the 1181 guest star, it is younger,
but physically larger and expanding more slowly than the Crab Nebula. Since the
kinematics of the ejecta in 3C~58 show no deceleration by the CSM found within
the remnant (shown by the lack of intermediate velocities), the slow expansion
velocities which characterize this remnant are difficult to explain. A
solution is simply to discard the remnant's proposed association with the 1181
event and assume 3C~58 is much older than 1000 yrs. Such a conclusion is
supported by several recent studies of the remnant. The synchrotron expansion
rate \citep{Biet06}, current internal energy of the PWN
\citep{Chevalier04,Chevalier05}, pulsar spindown age \citep{Murray02}, and the
amount of mass swept up by the PWN (as measured in the X-ray)
\citep{Chevalier04,Chevalier05} all suggest an age of a few thousand years. 



\end{document}